\documentstyle[twoside,11pt,paspconf,epsf]{article}
\def\etal   {{et~al.\/}}

\begin{document}
\title{Abundance Profiles in Low-mass Galaxies}
\author{Chip Kobulnicky}
\affil{UCSC/Lick Observatory \\ Santa Cruz, CA 95064 } 

\begin{abstract}
The nitrogen and oxygen abundances in the warm ionized gas of low-mass,
metal-poor galaxies appear surprisingly homogeneous considering the
prevalence of large HII regions, which contain hundreds
of massive stars.  Of the six galaxies with extensive optical
spectroscopy, only the largest and most metal-rich, the LMC, shows
evidence for a chemical gradient akin to those commonly seen in spirals.
Furthermore, no significant localized chemical fluctuations are found in the
vicinity of young star clusters, despite large expected chemical yields
of massive stars.  An ad-hoc fine-tuning of the release, dispersal
and mixing rates could give rise to the observed homogeneity, but 
a more probable explanation is that fresh ejecta reside in a
hard-to-observe hot or cold phase.  
In any case, the observations indicate that heavy
elements which have already {\it mixed} with the warm interstellar
medium are homogeneously {\it dispersed}.  Mixing of fresh ejecta with
the surrounding warm ISM apparently requires longer than the lifetimes of
typical HII regions ($>$10$^7$ yrs).  The lack of observed localized
chemical enrichments is consistent with a scenario whereby
freshly-synthesized metals are expelled into the halos of galaxies in a
hot, 10$^6$ K phase by supernova-driven winds before they cool and
``rain'' back down upon the galaxy, creating gradual enrichments on
spatial scales $>$1 kpc.

\end{abstract}


\section{Introduction}

Defining a sample of ``low-mass'' galaxies is immediately problematic
because very few systems, especially those of low mass,  have
well-determined dynamical masses.  Fortunately, optical luminosity
correlates well with the dynamical mass inferred from
galactic rotation curves (e.g., Tully \& Fisher
1977), and it is a reasonable substitute parameter.  
For purposes of this review, I
have taken ``low-mass'' to mean systems with $M_B$ fainter than $-18$.
The luminosity criterion for a ``dwarf'' galaxy is variously taken to
be $M_B>-18$ (Thuan 1983) or $M_B>-16$ (Tammann 1980; Hodge 1971).

Low-mass galaxies encompass a wide variety of nomenclature, including
blue compact dwarf galaxies (BCDG), HII galaxies, dwarf elliptical galaxies (dE), irregular, and low surface brightness (LSB) galaxies.  The chemical
composition of their stars and gas may be measured by optical
spectroscopy of emission lines from prominent HII regions (for warm
ionized gas), absorption lines in stellar atmospheres (for stars), or
X-ray spectroscopy of emission lines from highly ionized atomic species
(for hot diffuse gas).  Here I consider only abundance measurements in
the warm photo-ionized gas in and around HII regions.  Since oxygen is
the most easily measured element in HII regions, the terms
``abundance'' or ``metallicity'' should be read as the ``abundance of
oxygen relative to hydrogen.'' Reviews of the abundances of other
elements in stars (usually Fe) or hot coronal gas may be found
elsewhere in this volume.

The topic of abundance profiles entails two distinct issues.  
1) Do low-mass galaxies show {\it global} radial abundance gradients 
as large spiral galaxies (Searle 1971; Zaritsky, Kennicutt \& Huchra 1994) 
often do?  2) Are there {\it localized} chemical fluctuations
on spatial scales comparable to individual giant HII regions that might
be due to heavy elements recently synthesized and released by
massive stars (e.g., ``self-enrichment''---Kunth \& Sargent 1986;
``local contamination'---Pagel, Terlevich, \& Melnick1986).  Since chemical abundances
are like fossils that record the previous star formation activity, both
types of elemental variations contain information about
the star formation and gas dynamical history of the host galaxy.

\section{Radial Abundance Gradients}
\subsection{Object Selection}

Reliable measurement of a chemical spatial profile requires knowledge
of the chemical abundances spanning a range of radial distances.  To
date, the literature contains only a handful of low-mass galaxies with
optical spectroscopic abundance determinations at five or more distinct
locations.  The nine such galaxies are summarized in Table~1, along
with basic physical parameters necessary to interpret their abundance
profiles.  Most objects are relatively nearby and have Magellanic spiral or
irregular morphology.

\begin{table}
\caption{Low-Mass Galaxies with Abundance Measurements} \label{tbl-1}
\begin{center}
\begin{tabular}{llrrrr}
Name     & Morph  & Dist.    & M$_B$      & 12+log\tablenotemark{a} &  $h_0$\tablenotemark{b}           \\
         &        & (Mpc)    &            &  (O/H)                  & (kpc)         \\
\tableline                                      
NGC~6822 & IBm    & 0.47\tablenotemark{c} & -14.88\tablenotemark{c} & 8.19\tablenotemark{d}    & -- \tablenotemark{e}   \\
NGC~2366 & SBm    & 2.83\tablenotemark{c} & -15.79\tablenotemark{c} & 7.92\tablenotemark{f}    & 2.1\tablenotemark{g}   \\
NGC~1569 & IBm    & 1.72\tablenotemark{c} & -16.34\tablenotemark{c} & 8.19\tablenotemark{h}    & 0.3\tablenotemark{g}   \\
SMC      & Im     & 0.058\tablenotemark{c}& -16.35\tablenotemark{c} & 8.03\tablenotemark{i}    & 0.8\tablenotemark{j}   \\
F365-1   & Sm/LSB & 45\tablenotemark{k}   & -17.3\tablenotemark{k}  & 8.0\tablenotemark{k}     & 4.3\tablenotemark{k}   \\
UGC~5999 & Im/LSB & 45\tablenotemark{k}   & -17.7\tablenotemark{l}  & 8.0\tablenotemark{k}     & 4.3\tablenotemark{m}   \\
UGC~5005 & Im/LSB & 52\tablenotemark{k}   & -17.9\tablenotemark{l}  & 8.0\tablenotemark{k}     & 4.4\tablenotemark{m}   \\
LMC      & SBm    & 0.047\tablenotemark{c}& -17.73\tablenotemark{c} & 8.03\tablenotemark{i}    & 1.9\tablenotemark{j}   \\
NGC~4214 & SBm    & 4.03\tablenotemark{c} & -17.82\tablenotemark{c} & 8.19\tablenotemark{n}    & 1.7\tablenotemark{g}   \\
\tableline
\end{tabular}
\end{center}
\tablenotetext{a}{Oxygen abundances, 12+log(O/H) are nominal mean values for
general reference only, neglecting the possibility of internal
variations.} 
\tablenotetext{b}{Scale lengths, $h_0$, in kpc are derived
using the angular scale length of the B-band photometry from the listed
reference in conjunction with the assumed distance in column 3.}
\tablenotetext{c}{From the self-consistent distance and magnitude tabulation 
	of Richer \& McCall (1995).  See refs. therein.   }
\tablenotetext{d}{Pagel, Edmunds, \& Smith (1980). }
\tablenotetext{e}{No reliable published value. }
\tablenotetext{f}{Peimbert, Pena, \&  Torres-Peimbert (1986); see also Roy 
  \etal\ (1996)}
\tablenotetext{g}{de Vaucouleurs \etal\ (1991); 
  The angular disk scale length, $h_0$, is derived from half light radius,
	R$_d$, assuming an exponential disk so that R$_d$=1.678$h_0$.   }
\tablenotetext{h}{Kobulnicky \& Skillman (1997). }
\tablenotetext{i}{Pagel \etal\ (1978); for the LMC region N4A, the Pagel \etal\
value of 12+log(O/H)=8.44 has been replaced with a more recent measurement by
8.18 by Russel \& Dopita (1990). }
\tablenotetext{j}{Bothun \& Thompson (1988). }
\tablenotetext{k}{de Blok (1997); assumes H$_0$=75. }
\tablenotetext{l}{de Blok (1997); estimated from R magnitude assuming
	M$_B$$\approx$M$_R$+0.7 as is typical of other LSB galaxies in the 
	sample. }
\tablenotetext{m}{de Blok (1997);  scale length given is that for R band instead of B
	band. }
\tablenotetext{n}{Kobulnicky \& Skillman (1996).  }

\end{table}

\subsection{Oxygen Abundance Profiles}
Figure 1 shows the oxygen abundance as a function of radius for the objects in
Table~1 using the published galactocentric\footnote{deprojected for inclination
in all cases except NGC 1569} radial distance and O/H
measurements computed in the original references.  The
three LSB galaxies could not be shown because neither linear nor
angular radial distances are given in the original reference.  A single
error bar is plotted to provide an estimate of the typical 1$\sigma$ uncertainty
when O/H is computed empirically from [O~II], [O~III] and H$\beta$ line
ratios rather than from a direct determination of the electron
temperature via the [O~III] $\lambda$4363 line.  For objects such as
NGC~6822 and NGC~1569 which have O/H measurements using both methods,
results from the direct method are shown with large symbols and results
from empirical methods are shown with smaller symbols of the same type.
Note that there is typically a systematic offset of 0.1---0.2 dex between the two
methods, underscoring the need for direct O abundance determinations.

In Figure~1 there is no significant correlation of radial distance with
oxygen abundance except in the LMC where the gradient is
$-$0.048$\pm$0.019 dex/kpc.  In all other cases the O/H gradient is 
less than 0.02 dex/kpc, formally consistent
with zero.   However, only the LMC, SMC, and NGC~2366 have 
abundance measurements over a large radial extent.  High
quality spectra of HII regions in the outer parts of other irregular
galaxies will be necessary to detect small amplitude gradients,
if they exist.  

\begin{figure}
\plotfiddle{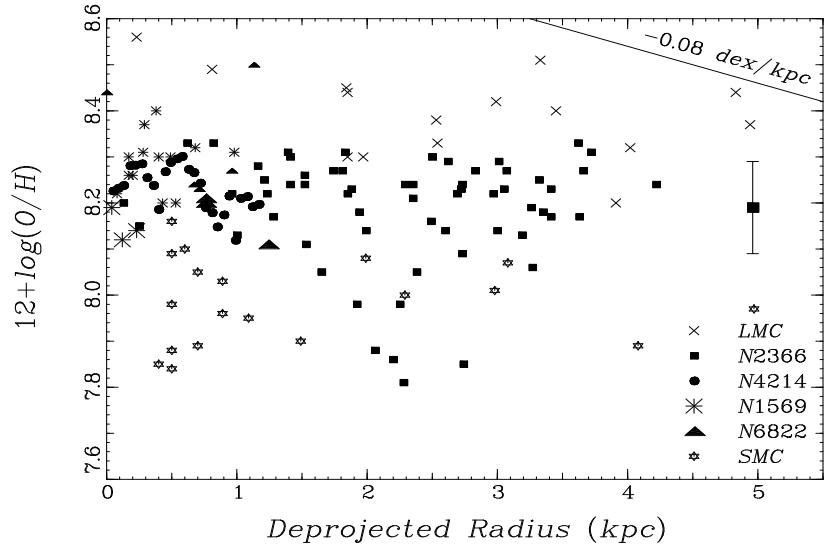}{65truemm}{-90}{50}{45}{-170}{250}
\caption{O/H versus deprojected radial distance in kpc for HII regions in
low-mass galaxies.  For all objects except the LMC, the slope of the O/H
gradient is consistent with 0.00$\pm$0.02 dex/kpc.  However, more high quality spectra of HII regions at large radii are necessary to establish
whether small gradients may exist. }
\end{figure}

In large spiral galaxies there has been considerable discussion about
which size parameter is the appropriate one to use in measuring radial profiles.  Many authors have chosen the
optical disk scale length, $h_0$, (see discussions in Zaritsky, Kennicutt,
\& Huchra, 1994; Garnett \etal\ 1997) in lieu of a fixed linear size.
The O/H ratio in low-mass galaxies as a function of optical scale
length is shown in Figure~2.  Only NGC~6822 is not plotted because it
does not have a published value for $h_0$.  The interpretation is
the same as Figure~1, in that the LMC is the only object with a
significant abundance gradient, $-0.097\pm0.037$ dex/$h_0$.
A robust limit on the gradients in the other systems will require
high-quality spectroscopy of HII regions beyond 2 optical scale lengths.

\begin{figure}
\plotfiddle{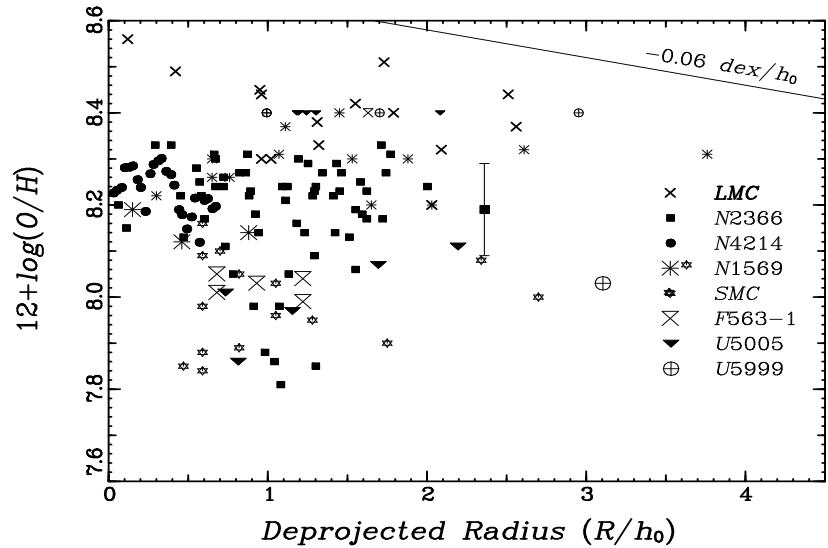}{65truemm}{-90}{50}{45}{-170}{250}
\caption{O/H versus deprojected radial distance in optical (B-band)
scale lengths for HII regions in
low-mass galaxies.  The figure is qualitatively similar to Figure~1,
with the addition of 3 LSB galaxies from de Blok (1997). Only
the LMC shows a significant radial oxygen gradient.  }
\end{figure}

\section{Localized Abundance Fluctuations}

\subsection{Do Chemical Fluctuations Exist?}

Clusters of massive stars are capable of creating large localized
chemical enrichments (Esteban \& Peimbert 1995), yet in the vicinity of
young starbursts in NGC~4214 (Kobulnicky \& Skillman 1996) and NGC~1569
(Kobulnicky \& Skillman 1997), no sizable O, N, or He anomalies are
seen in the surrounding warm photoionized medium.  NGC~5253 (Welch
1970; Walsh \& Roy 1989; Kobulnicky \etal\ 1997) remains the only
well-established
exception, containing central starburst region overabundant in nitrogen
by a factor of 3 compared to the surrounding ISM.  Further evidence for
chemical homogeneity in the ISM around starbursts is provided by the
two HII regions in the very metal-poor I~Zw~18.  Separated by $\sim$50
pc, they show identical O and N abundances (Skillman \& Kennicutt
1993; Martin 1996; Vilchez \etal\ 1997), even though the yield from a single massive star would be
sufficient to measureably ``pollute'' either one.

A quantitative comparison of expected chemical pollution 
versus observed abundance fluctuations is shown in Figure~3
for the case of the super star cluster A in NGC~1569.
The measured N/O ratio along a 45 arcsec strip adjoining 
the star cluster is plotted.  The N/O ratio is a particularly 
robust measure of potential abundance fluctuations because it
is relatively insensitive to errors in the adopted
electron temperature.  In Figure~3, no substantial variations beyond the
measurement uncertainties are evident  despite the sensitivity to
the N or O yields of just a few massive stars.
For example, the slightly-elevated N/O ratio seen at the
position number 12 at the 35$^{\prime\prime}$ mark could be caused by as few as two 
 40 M$_\odot$ stars.  

\begin{figure}
\plotfiddle{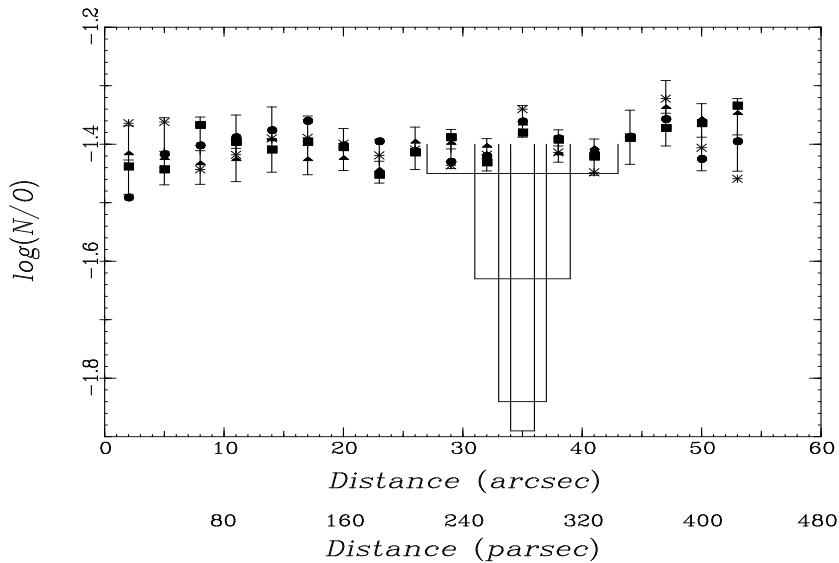}{65truemm}{-90}{50}{40}{-170}{225}
\caption{The N/O ratios along a 45$^{\prime\prime}$ strip of the ISM adjoining
cluster A in NGC~1569.  The expected magnitude
of chemical enrichment (predominantly O enrichment) is shown
by solid lines for four different dispersal radii.  The observed
variations are small in comparison to predicted enrichments, suggesting
some of the freshly-released elements are hidden in a hard-to-observe phase.}
\end{figure}

\subsection{Is Dilution the Solution to Pollution?}

Can the lack of observed enrichment be just due to rapid dispersal and
dilution of the heavy elements?  Not likely.  Assuming an age of 10 Myr
for cluster A, homogeneous dispersal within a sphere of a given radius,
a filling factor for the ionized gas of 0.1, a gas density of 100
cm$^{-1}$, an IMF slope of -2.7 in the range 0.5---100 M$_\odot$ (see
Kobulnicky \& Skillman 1997 for details), the expected N/O deviations
for a variety of dispersal scales are shown with solid lines.  Adopting
any combination of inhomogeneous dispersal, higher cluster age, lower
filling factor or lower average gas density will enhance the
expected chemical variations.  Given typical expansion speeds of
supernova ejecta, the heavy elements {\it could } be dispersed through
a larger region than the largest simulated volume, and thereby become
undetectable.  However, such an extreme rapid--dilution scenario
requires a finely--tuned, ad hoc dispersal mechanism to maintain the
appearance of homogeneity of scales of several hundred pc.  It seems
that the heavy elements produced by the massive stars (down to 20
M$_\odot$) in cluster A are evidently not seen in the warm ionized
medium.  

Kunth \& Sargent (1986) argued that objects more metal-poor than
I~Zw~18 should not be observed due to heavy element contamination from
the current burst of star formation.  In a small metal-poor galaxy like
I~Zw~18, a small number of O stars can produce significant chemical
enrichment.   Figure~4 sketches crudely the expected oxygen enhancement
in a hypothetical primordial--abundance galaxy as a function of oxygen
yield, gas mass/volume, and $nff$, the product of the volume--averaged
gas density and filling factor.  It helps illustrate that
only a few solar masses of O are required to raise the 
oxygen abundance from primordial to that seen in I~Zw~18
provided that only the inner few tens of pc are enriched.
However, longslit spectroscopic surveys suggest that the O
abundance is similar over several hundred pc, so perhaps
several thousand O stars would be required to raise the abundance to the
observed level.

\begin{figure}
\plotfiddle{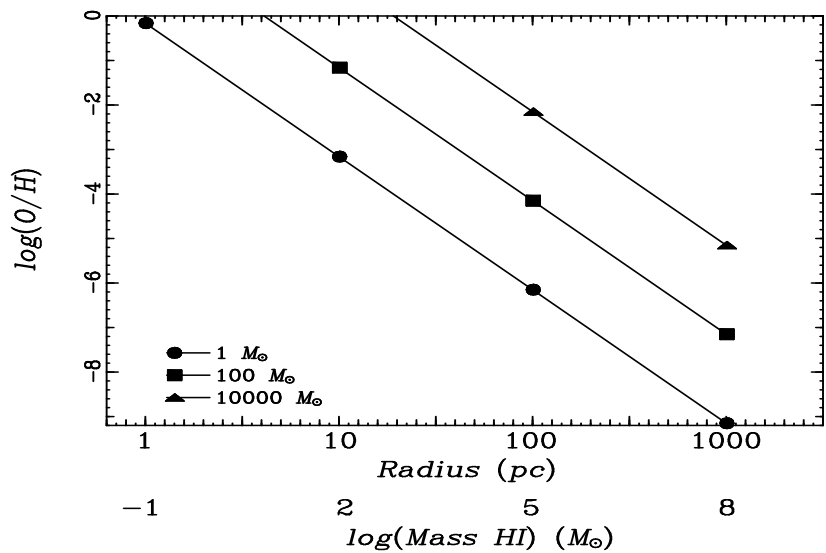}{65truemm}{-90}{50}{40}{-170}{230}
\caption{The resulting O/H ratio in a volume of gas initially
of primordial composition assuming homogeneous dispersal
of heavy elements throughout a sphere of a given radius.
The lower x-axis labels show the approximate hydrogen gas mass 
enclosed in the given volume assuming $nff$, the product of
mean gas density and filling factor is 1.0.  The resulting O/H 
ratio is shown for yields of 1 M$_\odot$, 100 M$_\odot$,
and 10,000 M$_\odot$ of oxygen.  }
\end{figure}

\subsection{Where Have all the Metals Gone?}

Given the lack of visible chemical enrichment around major starbursts,
several explanations could be considered.  {\it 1) Perhaps the metals were
never produced/released in the first place?}  If the number of massive
stars originally present in the cluster has been overestimated based on
the remaining stellar content, then the expected mass of heavy elements
would be reduced.  An abnormally low upper mass cutoff in the IMF, or
an abnormally steep IMF could work to accomplish the observed effect.
Yet, NGC~4214 and NGC~1569 do contain Wolf-Rayet stars, so clearly the
most massive stars have been, and are still present in the bursts.  

{\it 2) An
alternative suggestion requires that black holes left by supernovae from
massive stars to engulf the metals produced by the progenitor (e.g., Maeder
1992).}  This idea merits further theoretical investigation, but unless
the lower mass limit for black hole formation, M$_{BH}$ is considerably
lower than $\sim$50 M$_\odot$, then the reduction of chemical yields
will be too minor to resolve this problem.  See discussion elsewhere in
this volume for current estimates of M$_{BH}$.

{\it 3) The last, and most probable explanation for missing metals requires
that the freshly-ejected metals reside in a hard-to-observe hot or cold
phase (e.g., Tenorio-Tagle 1996).}  Since supernovae and the superbubbles formed by concerted
supernovae contain copious X-ray emitting material, hot gas is the
preferred explanation.  The Cas A supernova remnant, for example,
contains between 4 M$_\odot$ (Vink, Kaastra, \& Bleeker 1996) and 15 M$_\odot$ 
(Jansen, Smith, \& Bleeker 1989) of X-ray emitting material, consistent with the amount of
ejecta expected from the progenitor.  The next generation of orbitting
spacecraft should be able to measure the mass and metallicity of hot
gas surrounding massive star clusters and make a direct comparison to
expectations based on starburst models.



\acknowledgments 
I am indebted to Evan Skillman for collaboration
and comments on much of this work.  I also thank Jean-Ren\'e Roy
and the organizing committee for the invitation to present
this review.


\end{document}